\documentclass[12pt]{article}
\pdfoutput=1

\usepackage[small]{titlesec}
\usepackage{amsmath,amssymb,tikz-cd,graphicx,hyperref,physics}
\numberwithin{equation}{section}
\hypersetup{
colorlinks=false
}

\def\o{\omega}
\def\d{\mathrm{d}}
\def\be{\begin{equation}}
\def\ee{\end{equation}}

\begin{document} 
\title{Supertranslation Goldstone and de Sitter Tachyons}

\author{Hongjie Chen\footnote{hongjie.chen@pku.edu.cn}}
\date{}
\maketitle

\begin{center}
	{\it
		School of Physics, Peking University, No.5 Yiheyuan Rd, \\ Beijing 100871, P.~R.~China\\
	}
	\vspace{1cm}
\end{center}

\abstract{Supertranslation Goldstone lies in certain ``exceptional series'' representations of $SL(2,\mathbb{C})$. Interestingly, $m^2=-3$ scalar tachyon in three dimensional de Sitter space also lies in the same representation. In this note, we analyze these theories, focusing on representation-theoretical aspects, and emphasize that ``modulo certain polynomials'', there is a unitary representation of the corresponding symmetry group.}

\maketitle
\flushbottom

\section{Introduction}

In unitary Lorentzian conformal field theories, one deals with highest weight representations of $SO(d,2)$, where $d$ is the dimension of spacetime. In certain exotic CFTs, such as celestial CFT \cite{Pasterski:2016qvg, Pasterski:2017kqt,Pasterski:2017ylz,Strominger:2017zoo,Raclariu:2021zjz,Pasterski:2021rjz}, or the dS/CFT\cite{Strominger:2001pn}, since the symmetry group is Lorentz group $SO(1,d+1)$, fields or particles are organized into unitary representation of $SO(1,d+1)$.

In the context of celestial holography, one Mellin transforms on energy, turning momentum eigenstates into boost eigenstates, and scattering amplitudes into two dimensional correlation functions.
The boost eigenstates are labelled by their conformal dimension $\Delta$ and spin $s$. A complete basis of states is given by those on the principal series $\Delta=1+i\delta$, where $\delta$ is real.
However, those related to soft theorems \cite{Weinberg:1965nx} and generate asymptotic symmetries \cite{Strominger:2013lka,Strominger:2013jfa,He:2014laa,He:2014cra}, have integer $\Delta=1,0,-1,\cdots$ \cite{Donnay:2018neh}. These are called \textit{conformally soft} sector.

In \cite{Donnay:2020guq}, it was argued that these integer valued conformal dimensions can be understood as analytic continuation in $\Delta$. There are intertwining relations between representations with different conformal weights and spins. The structure is called ``celestial diamond'' \cite{Pasterski:2021fjn,Pasterski:2021dqe}. It is similar to the discussion of global primary descendants occurring at (half-)integer values of conformal weight in Lorentzian CFTs \cite{Penedones:2015aga}. Using celestial diamonds, we can build irreducible (and possibly, unitary) representations out of each diamond. A surprising fact is that many of these representation are equivalent. Goldstones and currents lie in these irreducible representations. 

Supertranslation Goldstone is related to the leading order soft theorem for gravitons. One might wonder whether they furnish a unitary representation of the symmetry group $SL(2,\mathbb{C})$, since the bulk theory is unitary. The answer is yes, but one need to quotient out the polynomials corresponding to rigid Poincar\'e transformations. This is seen through its effective action $S\propto \int d^2 z C(z,\bar z)\square^2 C(z,\bar z)$.
Since near spacelike infinity four dimensional de Sitter space is can be foliated into slices of three dimensional de Sitter space, it is also expected to relate to certain scalar theory in three dimensional de Sitter space. This is also true, and corresponding scalar has mass squared $m^2=-3$.

 The $\square^2$ theory belong to a series of more general scalar theories in two dimensions $S\propto \int d^2 z C(z,\bar z)\square^k C(z,\bar z)$, where $k\ge 1$. The case $d=2, k=1$ is just the usual free scalar, which enjoys a larger symmetry group, Virasoro group. It is the effective action of Goldstone corresponding to leading soft theorem for photons.
For $k>1$, there is only global conformal symmetry group $SL(2,\mathbb{C})$.

The $m^2=-3$ scalar in three dimensional de Sitter space also belongs  to a larger class called ``exceptional scalars'' in $d+1$-dimensional de Sitter space \cite{Bros:2010wa,Epstein:2014jaa,Bonifacio:2018zex,Bonifacio:2021mrf}, with mass squared 
\be
\label{mass2}
    m^2=-(s-1)(s-1+d),\; s\in\mathbb{Z}_+\, .
\ee
The case relevant for supertranslation Goldstone is $d=2,s=2$.

This note is organized as follows. In section \ref{4}, we review the supertranslation, celestial basis and ``celestial diamonds''. In section \ref{sl2} and \ref{rep}, we review the representation theory of $SL(2,\mathbb{C})$ and $SO(1,d+1)$ relevant to this note. In section \ref{2}, we analyze the effective action for supertranslation Goldstones, we point out if modulo the polynomials corresponding to rigid translation of Poincar\'e group, we can get a unitary representation of $SL(2,\mathbb{C})$. In section \ref{3}, we generalize the story above and go to $d+1$ dimensional de Sitter space, we point out for scalars with mass squared (\ref{mass2}), modulo some pseudo-harmonic polynomials in $\mathbb{M}^{1,d+1}$, we can get a unitary representation of $SO(1,d+1)$.

\section{Supertranslation Goldstones}\label{4}
In this section we review supertranslation Goldstone modes, the notion of conformally soft operators, and the celestial diamond for supertranslation Goldstone.

\subsection{The BMS analysis}

Supertranslations are asymptotic symmetries of asymptotically flat space in four dimensions \cite{Bondi:1962px,Sachs:1962wk}, which are large diffeomorphisms near lightlike infinity. Under supertranslation, one spacetime turns into another, and hence supertranslation symmetry is spontaneously broken if one chooses one particular spacetime.  Supertranslation symmetry is related to leading soft graviton theorem \cite{Strominger:2013jfa,He:2014laa}. The supertranslation Goldstone is related to soft factorization and gravitational dressing \cite{Himwich:2020rro,Arkani-Hamed:2020gyp,Pasterski:2021dqe}.

Using retarded coordinates
\begin{align}
ds^2&=-du^2-2dudr+2r^2\gamma_{z\bar z}dzd\bar z\\
&\frac{2m_B}{r}du^2+r C_{z\bar z} dz^2+rC_{\bar z\bar z}d\bar z^2+D^zC_{zz}dudz+D^{\bar z}C_{\bar z\bar z}dud\bar z+\cdots,
\end{align}
supertranslation is generated by vector field near null infinity
\be
\zeta=f\partial_u-\frac{1}{r}(D^z f\partial_z+D^{\bar z}f\partial_{\bar z})+D^{z}D_z f\partial_r+\cdots,
\ee
where $f=f(z,\bar z)$, and the corresponding transformations of $C_{zz}$ is
\be
\mathcal{L}_{f} C_{zz}=f\partial_u C_{zz}-2D_z^2 f.
\ee
If one defines \cite{He:2014laa}
\be
C_{zz}|_{\mathcal{I}^+_-}=-2D_z^2 C,
\ee
then
\be
\mathcal{L}_{f} C=f.
\ee
Thus $C$ is the Goldstone boson of spontaneously broken supertranslation symmetry.

Write the massless momentum as
\be
k^{\mu}=\omega q^\mu=\omega(1+z\bar z,z+\bar z, -i(z-\bar z),1-z\bar z).
\ee
Massless particles transform under supertranslation as \cite{Himwich:2020rro}
\be
\delta_f  O(\eta,\omega,z,\bar z)=i\eta \omega f(z)O(\omega,z,\bar z)
\ee
where $\eta=\pm 1$ denotes outgoing or incoming.
Define the operator
\be
W(\eta,\omega,z,\bar z)=e^{i\eta\omega C(z,\bar z)},\; O=W\tilde O,
\ee
which absorbs the supertranslation transformation of $O$.

The soft factorization of a massless scattering amplitude can be made explicit by this new operator $W$
\be
\langle O_1(k_1)\cdots O_n(k_n)\rangle=\langle W_1(k_1)\cdots W_n(k_n)\rangle\langle \tilde O_1(k_1)\cdots \tilde O_n(k_n)\rangle.
\ee
Comparing with the result of soft factorization \cite{Naculich:2011ry}
\be
\mathcal{A}_{soft}=\exp\left[ \frac{G}{2\pi\epsilon}\sum_{i\ne j}k_i\cdot k_j \ln\left(\frac{2k_i k_j}{\mu^2}\right)  \right],
\ee
the correlation function of $C(z,\bar z)$ is
\be\label{ctwo}
\langle C(z,\bar z)C(0,0)\rangle=\frac{2G}{\pi \epsilon}|z|^2\ln |z|^2.
\ee

In \cite{Nguyen:2021ydb}, this two point function was derived through effective action of supertranslation Goldstones
\be
S\propto\int d^2 z C(z,\bar z)\square^2 C(z,\bar z).
\ee
The effective action was derived using the hyperbolic slicing and the action in \cite{Compere:2011ve} at spacelike infinity. In fact, since the two point function is purely kinematical, it (along with the free action) can be derived directly from the conformal dimension of the supertranslation Goldstone, which we now review.

\subsection{Conformal primaries and conformally soft sector}

For massless field $\phi_s(X)$ with spin $s$, conformal primary operators can be extracted from bulk field operator
\be
\label{primary}
\mathcal{O}^{\pm}_{\Delta,J}(q) = i\left( \phi_s (X), \Phi^{s}_{\Delta^*,-J}(X_{\mp},q)\right)_\Sigma,
\ee
where $\pm$ denotes in or out state, $J=\pm s$ is the helicity, and $X_{\pm}=X\pm i(1,0,0,0)$. The right hand side is suitable inner product for this field on a Cauchy slice $\Sigma$. The conformal primary wavefunction $\Phi^{\pm}_{\Delta,J}(X,q)$  transforms under Lorentz group $SO(1,3)$ as \cite{Pasterski:2017kqt}
\be
\Phi^s_{\Delta,J}\left(\Lambda^\mu{}_\nu X^\nu;\frac{az+b}{cz+d},\frac{\bar a\bar z+\bar b}{\bar c\bar z+\bar d}\right)=(cz+d)^{\Delta+J}(\bar c\bar z+\bar d)^{\Delta-J}D_s(\Lambda)\Phi^s_{\Delta,J}(X^\mu;z,\bar z).
\ee
The conformal primary wavefunctions can be obtained by Mellin transformation on energy $\omega$ of the plane wave solutions, so (\ref{primary}) is in fact a Mellin transformed version of the more familiar bulk equation of expressing annihilation operator as inner product between the field and plane wave basis.

Soft theorems are related to poles and zeros of energy $\omega$, and Mellin transformation turns it into poles in $\Delta$  \cite{Pate:2019mfs,Arkani-Hamed:2020gyp}. Consider $n+1$ massless particles scattering, suppose we take the energy of one particle to be soft $k^\mu=\o q^\mu, \omega\rightarrow 0$. At tree level, the $\mathcal{S}$-matrix element develops a Laurent expansion in $\o$,
\be
A_{n+1}|_{\omega\rightarrow 0}=\omega^{-1}A^{(-1)}+A^{(0)}+\omega A^{(1)}+\cdots.
\ee
Due to
\be
\int_0^\Lambda \d \o \o^{\Delta-1}\o^n=\frac{\Lambda^{\Delta+n}}{\Delta+n},
\ee
the Mellin transformed amplitude has poles in $\Delta=1,0,-1,\cdots$, and soft theorems have a two dimensional interpretation as insertion of operators with dimensions $\Delta=1,0,-1,\cdots$ on the celestial sphere. These operators are called conformally soft operators.

The leading order soft theorem for gravitons corresponds to a current operator with $\Delta=1,J=\pm 2$. The corresponding Goldstone current also has $\Delta=1,J=\pm 2$. The conformal primary wave functions of them are symplectic partners of each other \cite{Donnay:2018neh}. As observed in \cite{Pasterski:2021fjn,Pasterski:2021dqe}. They are part of the ``celestial diamonds''. For Goldstone, the diamond reads
\be
\label{diamond}
\begin{tikzcd}
& C\arrow[dl,"\partial^2"]\arrow[dr,"\bar\partial^2"] &\\
C_{zz}\arrow[dr,"\bar\partial^2"]&&C_{\bar z\bar z}\arrow[dl,"\partial^2"]\\
&C_{zz\bar z\bar z}&
\end{tikzcd}
\ee
where $C_{zz},\, C_{\bar z\bar z}$ are the Goldstones currents with $\Delta=1,\, J=\pm 2$.
The top term is a conformal primary with $\Delta=-1, J=0$, which is exactly the supertranslation Goldstone $C$ introduced in the previous section.

The articles \cite{Pasterski:2021fjn,Pasterski:2021dqe} dealt with representations of the conformal algebra. But the discussion is readily promoted to the whole $SL(2,\mathbb{C})$ group. In fact, there is a unitary representation of $SL(2,\mathbb{C})$ underlying the diagram (\ref{diamond}), as  reviewed in section \ref{sl2}.

\section {Exceptional representations of $SL(2,\mathbb{C})$}\label{sl2}
There is a two-to-one homomorphism
\be
SL(2,\mathbb{C})\rightarrow SO^+(3,1),
\ee
so we consider representations of $SL(2,\mathbb{C})$.

In the section, we review  ``integer points'' representations of $SL(2,\mathbb{C})$. The study of unitary representations of Lorentz group was initiated in classic works \cite{harish1947infinite,bargmann1947irreducible,gel1947unitary}. We follow Chapter {\MakeUppercase{\romannumeral 3}} of \cite{gelfand1966generalized}.

We use $D_{(n_1,n_2)}$ to denote the space of functions on the complex plane $z,\bar z$, which transform under 
\be
\begin{pmatrix}
a&b\\
c&d 
\end{pmatrix}\cdot z=\frac{az+b}{cz+d}=w
\ee 
as
\be
\phi'(z,\bar z)=(cz+d)^{n_1-1}(\bar c \bar z+\bar d)^{n_2-1}\phi(w,\bar w),
\ee
and satisfies boundary condition
\be
\label{zasymp}
\phi(z,\bar z)\sim C z^{n_1-1}\bar z^{n_2-1}, \;\mathrm{when} \; z\rightarrow \infty.
\ee
which means it grows no faster than $ z^{n_1-1}\bar z^{n_2-1}$.
We can also write $2h_{\phi}=1-n_1,\; 2\bar h_{\phi}=1-n_2$.

Note that this is dual of the transformation law of quasi-primary operator, 
\be
O'(w,\bar w)=(cz+d)^{2h_O}(\bar c\bar z+\bar d)^{2\bar h_O}O(z,\bar z),
\ee
in the sense that if we obtain states from acting $O(z,\bar z)$ on the vacuum,
\be
\ket{z,\bar z}=O(z,\bar z)\ket{0},
\ee
then transformation of the smeared state
\be
\ket{\phi}=\int d^2 z \phi(z,\bar z) \ket{z,\bar z}
\ee
can be either viewed as transforming $O(z,\bar z)$ with $\phi(z,\bar z)$ fixed 
\be
\ket{\phi'}=\int d^2 w \phi(w,\bar w) O'(w,\bar w)\ket{0},
\ee
or the other way round
\be
\ket{\phi'}=\int d^2 z \phi'(z,\bar z) O(z,\bar z)\ket{0}.
\ee
$h_\phi$ and $h_O$ are related by
\be
h_\phi+h_O=1,\; \bar h_\phi+\bar h_O=1.
\ee

We are interested in the case where $n_1,\;n_2$ are integers. There are intertwining relations between the representations
\begin{equation}
\label{sl2diagram}
\begin{tikzcd}
                                     &E_{(n_1,n_2)}\arrow[d,hook]        & 0\arrow[l]\arrow[d]      &\\
F_{(-n_1,-n_2)}\arrow[d] & D_{(n_1,n_2)} \arrow[l]\arrow[d,"\bar\partial^{n_2}"]\arrow[r,"\partial^{n_1}"]\arrow[dr,leftrightarrow] &D_{(-n_1,n_2)}\arrow[r]\arrow[d,"\bar\partial^{n_2}"]\arrow[dl,leftrightarrow] &0\arrow[d]\\
0\arrow[r]                      &D_{(n_1,-n_2)}\arrow[r,"\partial^{n_1}"]\arrow[d]                    & D_{(-n_1,-n_2)}\arrow[d]& F_{(-n_1,-n_2)}\arrow[l,hook]\\
&0&E_{(n_1,n_2)}\arrow[l]&
\end{tikzcd}
\end{equation}
In the figure, $n_1,\,n_2$ are positive. $E_{(n_1,n_2)}$ is spanned by polynomials in $z,\bar z$ of degree no higher than $n_1-1$ in $z_1$, and $n_2-1$ in $\bar z$. $F_{(-n_1,-n_2)}$ is spanned by functions $\phi$ which vanishes when integrated with the polynomials in $E_{(n_1,n_2)}$, i.e.,
\be
\int d^2 z\phi(z,\bar z)z^i\bar z^j=0,
\ee
with $i\le n_1-1,\,j\le n_2-1$.
The map from $D_{(-n_1,-n_2)}$ to $D_{(n_1,n_2)}$ is given by
\be
\phi'(z,\bar z)=\int d z_1 d\bar z_1 (z-z_1)^{n_1-1} (\bar z-\bar z_1)^{n_2-1} \phi(z_1,\bar z_1),
\ee
which is the shadow transformation. The map from $D_{(n_1,-n_2)}$ to $D_{(-n_1,n_2)}$ is given by
\be
\phi'(z,\bar z)=\int  d\bar z_1  (\bar z-\bar z_1)^{n_2-1} \partial^{n_1}\phi(z,\bar z_1),
\ee
and similarly for the map from $D_{(-n_1,n_2)}$ to $D_{(n_1,-n_2)}$.

All the sequences containing two and only two $D$ are exact. We have isomorphisms
\be
D_{(-n_1,n_2)}\cong D_{(n_1,-n_2)}\cong F_{(-n_1,-n_2)}\cong D_{(n_1,n_2)}/E_{(n_1,n_2)}.
\ee

For integer $n_1,\,n_2$, if and only if $n_1=n_2=n$, there is $SL(2,\mathbb{C})$ invariant positive definite hermitian inner product defined on $D_{(n,n)}/E_{(n,n)}$
\be\label{in1}
(\psi,\phi)=(-1)^n\int d^2 z \bar\psi(z,\bar z)\partial^n\bar\partial^n\phi(z,\bar z),
\ee
Through the diagram (\ref{sl2diagram}), the inner product is equivalent to that on 
on $D_{(-n,n)}$ or $D_{(n,-n)}$
\be\label{in2}
(\psi,\phi)= \int d^2 z \bar \psi(z,\bar z)\phi(z,\bar z),
\ee
and on $F_{(-n,-n)}$ subspace of $D_{(-n,-n)}$
\be\label{in3}
(\psi,\phi)=\int d^2 z |z_1-z_2|^{2n-2}\ln |z_1-z_2|^2\bar\psi(z_1,\bar z_1)\phi(z_2,\bar z_2).
\ee

\section{$\square^2$ scalar theory in two dimensions}\label{2}
To construct a free theory of scalar with dimension $\Delta=-1$ is simple. One considers the action \cite{Nguyen:2021ydb,Pasterski:2021dqe}
\be\label{action}
S=g\int d^2 z \partial\partial C(z,\bar z)\bar\partial\bar\partial C(z,\bar z)
\ee
of a real scalar field on Euclidean plane.

In two dimensions, we have $\square=4\partial\bar\partial$. We can also write the action as
\be
S=\frac{g}{16}\int d^2 z C(z,\bar z)\square^2 C(z,\bar z).
\ee

The theory is studied in, for example, \cite{wiese1996classification,Karananas:2015ioa}.
It does not have a local, symmetric, traceless stress tensor, and thus cannot couple to gravity \cite{Karananas:2015ioa,Brust:2016gjy} or have full Virasoro invariance. The reason is that the Paneitz operator \cite{paneitz2008quartic}, the Weyl covariant generalization of $\square^2$, diverges at dimension two. 

We can also see this fact by simpler methods. If we had a symmetric traceless stress tensor, we could write in the complex coordinates
\begin{align}
T_{z\bar z}=T_{\bar z z}&=0,\\
\partial T_{\bar z\bar z}&\approx0,\\
\bar \partial T_{ z z}&\approx0,
\end{align} 
where $\approx$ means on-shell equal to.
Now simply postulate $T_{z z}=\sum  \partial^{a_1}\bar\partial^{a_2}C \partial^{b_1}\bar\partial^{b_2}C$, with $a_1+b_1=3,\, a_2+b_2=1$. The only solution to $\bar\partial T_{ z z}\approx 0$ is zero. One might also try to postulate $T_{ z  z}=\sum  z^n\bar z^m \partial^{a_1}\bar\partial^{a_2}C \partial^{b_1}\bar\partial^{b_2}C$, with $-n+a_1+b_1=3,\, -m+a_2+b_2=1$. But the solution will involve infinite sum of $m$, making the stress tensor nonlocal.

The scaling dimension of $C$ is $-1$. The Green's function has the form
\be
\label{ln}
\langle C(z,\bar z)C(0)\rangle\propto |z|^2\ln |z|^2,
\ee
and can be positive or negative, similar to the case of free scalar in two dimensions
\be
S\propto \partial\phi\bar\partial\phi,
\ee
with $\langle \phi(z,\bar z)\phi(0)\rangle\propto \ln |z|^2$. In that case, there is a constant shift symmetry
\be
\phi\rightarrow \phi+c,
\ee
and one should consider the operators invariant under this symmetry, for example $\partial \phi$ or $\bar\partial\phi$. These two operators are conformal primary operators.

In the present case, we need to mod out symmetries generated by polynomials
\be\label{zshift}
C\rightarrow C+c_0+c_1 z+\bar c_1 \bar z+c_2 z\bar z.
\ee
These polynomials form $E_{(2,2)}$ in the notation of section \ref{sl2}.
Alternatively, if we view $C(z,\bar z)$ as operators acting on the ground state, we can define the smeared state
\be
\ket{\phi}=\int d^2 z \phi(z,\bar z)C(z,\bar z)\ket{0}=\int d^2 z \phi(z,\bar z)\ket{z,\bar z},
\ee
and let conformal generators act on $\phi(z,\bar z)$. The functions $\phi(z,\bar z)$ must satisfy
\be
\int d^2 z \phi(z,\bar z) (c_0+c_1 z+\bar c_1 \bar z+c_2 z\bar z)=0.
\ee
These $\phi(z,\bar z)$ form $F_{(-2,-2)}$ in the notation of section \ref{sl2}.

The invariant operator under (\ref{zshift}) constructed from $C$ must have at least two $\partial$ or two $\bar\partial$ acting on it. For example, $\partial^2C$ or $\bar\partial^2C$, or any derivatives of them.

One sees that
\begin{align}
\langle\partial^2C(z,\bar z)\partial^2C(0)\rangle&\propto\frac{\bar z}{z^3}\label{z},\\
\langle\bar\partial^2C(z,\bar z)\bar\partial^2C(0)\rangle&\propto\frac{\bar z}{z^3}\label{zbar},\\
\langle\partial^2C(z,\bar z)\bar\partial^2C(0)\rangle&\propto\delta(z) \label{deltaz},\\
\langle\partial^2\bar\partial^2C(z,\bar z)\partial^2\bar\partial^2C(0)\rangle&\propto\partial^2\bar\partial^2\delta(z).\label{delta2}
\end{align}
$\partial^2C(0)$ and $\bar\partial^2C(0)$ play the role of conformal primary operators. The representations they generate is isomorphic to each other, which is representation $D_{(-2,2)}\cong D_{(2,-2)}$ in notation of section \ref{sl2}. 
$\partial^2C(0)$ and $\bar\partial^2C(0)$ are hermitian conjugate to each other, because $z$ is genuinely complex. The two point function (\ref{deltaz}), leads to the hermitian inner product (\ref{in2}). The inner product is positive definite, which means we have a unitary representation of $SL(2,\mathbb{C})$.

As mentioned earlier, if insisting on considering $C(z,\bar z)$, we need to mod out four polynomials (\ref{zshift}). Thus although the two point function (\ref{ln}) does not lead to a positive definite inner product on the whole space $D_{(-2,-2)}$, it is positive definite on the subspace $F_{(-2,-2)}$. It is just (\ref{in3}).
Similarly, (\ref{delta2}) leads to positive definite inner product on $D_{(2,2)}/E_{(2,2)}$ (\ref{in1}).

Thus we have modeled the celestial diamond (\ref{diamond}) as a $\square^2$ scalar field theory. Modulo polynomials (\ref{zshift}), $\square^2$ scalar  describes a $SL(2,\mathbb{C})$ unitary theory.

Actually, the four real parameters (\ref{zshift}) have the bulk interpretation of the four rigid translations in Poinca\'e group. Doing a rigid translation does not change $C_{zz}$ or $C_{\bar z\bar z}$. The rigid translation symmetry is not broken. To consider physically distinct vacua related by supertranslation, one must mod out these rigid translations.

\section{Exceptional representation of $SO(1,d+1)$}\label{rep}
In this section, we review a special series of representation of generalized Lorentz group $SO(1,d+1)$ with $d\ge 2$ arising from ``integer points'' of elementary representations \cite{Dobrev:1977qv,Sun:2021thf}.

\subsection{Irreducible representations from exceptional series}
Irreducible representations can be produced from elementary ones $R_{\Delta,s}$ \cite{Dobrev:1977qv}, where $\Delta$ is the conformal weight, and $s$ labels the $SO(d)$ content. The representation space can be realized as functions on $\mathbb{R}^d$ or $S^d$. We choose $\mathbb{R}^d$ for now, and focus on the cases where $s$ is one-line Young diagram, which means it is a symmetric traceless $s$-tensor of $SO(d)$.

The functions $\psi_{\mu_1\cdots\mu_s}(x_i\cdots x_d)$ can be packaged into a compact form by contracting with a null vector $z$, where $z\in \mathbb{C}^d$ and $z^2=0$. Namely
\begin{equation}
    \psi(x,z)=\frac{1}{s^2}\psi_{\mu_1\cdots\mu_s}(x_i\cdots x_d)z^{\mu_1}\cdots z^{\mu_s}.
\end{equation}
The generators of $SO(1,d+1)$ act on the functions as follows \cite{Sun:2021thf}
\begin{align}
\label{p-action}
    P_i\psi(x,z)&=-\partial_i\psi(x,z),\\
    D\psi(x,z)&=-(x\cdot \partial x+\Delta)\psi(x,z),\\
    M_{ij}\psi(x,z)&=(x_i\partial_j-x_j\partial_i+z_i\partial_{z_j}-z_j\partial_{z_i})\psi(x,z),\\
    K_i\psi(x,z)&=\left(x^2\partial_i-2x_i(x\cdot\partial_x+\Delta)-2x^j(z_i \partial_{z_j}-z_j\partial_{z_i})\right)\psi(x,z).\label{k-action}
\end{align}
The asymptotic behavior of $\psi(x)$ is
\be
\psi(x)\sim\frac{C}{(x^2)^\Delta},
\ee
which means it grows no faster than $(x^2)^{-\Delta}$.

As in section \ref{sl2}, this transformation is dual of the transformation of local operators
\begin{equation}
\ket{x}_{\mu_1\cdots\mu_s}=O^{\bar\Delta}_{\mu_1\cdots\mu_s}(x)\ket{0}.
\end{equation}
with $\bar \Delta=d-\Delta$, and a general state is give by smearing
\begin{equation}
\ket{\psi}=\int d^d x\psi_{\mu_1\cdots\mu_s}(x) \ket{x}_{\mu_1\cdots\mu_s}.
\end{equation}
The action of conformal generators on $\ket{x}_{\mu_1\cdots\mu_s}$ is
\begin{align}
\label{xp-action}
    P_i \ket{x}_{\mu_1\cdots\mu_s}&=\partial_i\ket{x}_{\mu_1\cdots\mu_s},\\
    D\ket{x}_{\mu_1\cdots\mu_s}&=(x\cdot \partial x+\bar\Delta)\ket{x}_{\mu_1\cdots\mu_s},\\
    M_{ij}\ket{x}_{\mu_1\cdots\mu_s}&=(-x_i\partial_j+x_j\partial_i+\mathcal{M}^{(s)}_{ij})\ket{x}_{\mu_1\cdots\mu_s},\\
    K_i\ket{x}_{\mu_1\cdots\mu_s}&=\left(-x^2\partial_i+2x_i(x\cdot\partial_x+\bar\Delta)-2x^j\mathcal{M}^{(s)}_{ij}\right)\ket{x}_{\mu_1\cdots\mu_s}.\label{xk-action}
\end{align}
where $\mathcal{M}^{s}$ is the spin $s$ representation of $\mathfrak{so}(d)$. \footnote{Note that the convention here is different from those of 2d CFT in section \ref{sl2}.} The conformal generators acting on the smeared states $\ket{\psi}$ can be realized as transforming $ \ket{x}_{\mu_1\cdots\mu_s}$ while holding $\psi(x,z)$ fixed or the other way round.

We want to discuss the ``exceptional'' series, which have integer $\Delta$, 
 \begin{equation}
    R_{1-s,t}, R_{d+s-1,t}, R_{1-t,s}, R_{d+t-1,s}\, .
\end{equation}
Here $s, t$ are non-negative integers. We suppose that $s\ge t$. They all have the same Casmir $\mathcal{C}_2=-(s-1)(s+d-1)-t(t+d-2)$. Note that we do not use brackets in the subscript, to distinguish from notations in section \ref{sl2}. In the notations here, the first subscript denotes the conformal dimension, and the second denotes spin.

We use $C$ to denote the representation space. There are morphisms between these spaces, shown in the diagram below \cite{Dobrev:1977qv}
\begin{equation}
\label{diagram}
    \begin{tikzcd}
 0\arrow[r] \arrow[dr] & E_{s,t}\arrow[r,hook] & C_{1-s,t} \arrow[r, "S^+_{1-s,t}",shift left=1.5] \arrow[d, "d^{s-t}"]
    & C_{d+s-1,t} \arrow[l, "S^-_{1-s,t}"]  & F_{s,t} \arrow[l,hook] &  0\arrow[l]\arrow[dl]\\
& F'_{s,t}\arrow[r,hook] & C_{1-t,s} \arrow[r, "S^+_{1-t,s}",shift left=1.5]   & C_{d+t-1,s} \arrow[l, "S^-_{1-t,s}"] \arrow[u, "d'^{s-t}"] & D_{s,t} \arrow[l,hook] &
\end{tikzcd}
\end{equation}

The morphisms $S$ are shadow transforms, which relate representation $R_{\Delta,s}$ and  $R_{d-\Delta,s}$, while
\begin{equation}
    d^{s-t}=(z\cdot \partial_x)^{s-t},
    d'^{s-t}=(D_z\cdot \partial _x)^{s-t},
\end{equation}
where $D_{z^i}=\partial_{z^i}-\frac{1}{d+2(z\cdot \partial_z-1)} z_i \partial_z^2$ is the derivative which respects the null condition. The spaces $E_{s,t}, D_{s,t}, F_{s,t}, F'_{s,t}$ are respectively
\be
    E_{s,t}=\mathrm{ker}(d^{s-t}),\quad D_{s,t}=\mathrm{ker}(d'^{s-t}),\quad F_{s,t}=\mathrm{Im}(d'^{s-t}),\quad F'_{s,t}=\mathrm{Im}(d^{s-t}).
\ee
All the sequences in the diagram are exact. By using this fact, we can obtain irreducible representations
\begin{align}
E_{s,t}&\cong C_{d+s-1,t}/F_{s,t},\\
D_{s,t}&\cong C_{1-t,s}/F'_{s,t},\label{D=C/F}\\
V_{s,t}&\equiv C_{1-s,t}/E_{s,t}\cong F'_{s,t}\cong F_{s,t}\cong C_{d+t-1,s}/D_{s,t}.\label{repr-V}
\end{align}

$E_{s,t}$ is finite dimensional, thus cannot be unitary unless it is trivial. $D_{s,t}$ is unitary, and corresponds to partially massless fields in de Sitter space \cite{Deser:2001pe,Deser:2001us}. $V_{s,t}$ is unitary only when $t=0$.

\subsection {$E_{s,0}$ and $V_{s,0}$}
 Viewed as functions in $\mathbb{R}^d$,
$E_{s,0}$ is the linear span of functions $g(1,x_1,...,x_d, x^2)$, where $g(a_1,..,a_{d+2})$ is a homogeneous polynomial of degree $s-1$. For example, when $s=1$, $E_{s,0}$ is one-dimensional, the constant functions. For $s=2$, The functions are spanned by $x_1,...,x_d, x^2$ and constant. 

We need to prove that monomials of $x_1,...,x_d,x^2$ of degree $\le s-1$ ($x^2$ viewed as degree 1) are annihilated by $(z\partial_x)^s$. The idea is that $z\partial_x$ acting $s$ times on a monomial
\begin{equation}
    (x^2)^{n_0} x_1^{n_1}...x_d^{n_d},
\end{equation}
with $n_0+n_1+...+n_d\le s-1$, there must be one $x_i$ or $x^2$ on which $z\partial_x$ act at least twice. The result is zero, according to $s=2$ case. This argument also shows $E_{s,0}$ contains only these functions.

$E_{s,0}$ can also be viewed as the space of degree $s-1$ pseudo-harmonic polynomials in $d+2$ dimensional Minkowski space $\mathbb{M}^{1,d+1}$ (this means they are annihilated by the operator $\square=\partial^\mu\partial_\mu$), which are the polynomials that generate shift symmetries in (\ref{shift-symmetry}). This is easily seen by embedding $\mathbb{R}^d$ into the lightcone of $\mathbb{M}^{1,d+1}$,
\be
X^\mu=\lambda \left(\frac{1+x^2}{2},x^i,\frac{1-x^2}{2}\right),\, \lambda\in \mathbb{R}
\ee
where $1\le i\le d$ and $0\le \mu\le d+1$. Or, to relate to the next section, we first embed $d+1$ dimensional de Sitter space 
\begin{equation}
    ds^2=\frac{1}{\eta^2}(-d \eta^2+\sum_i dx_i^2)
\end{equation}
into $\mathbb{M}^{1,d+1}$,
\begin{align}
    X^0&=\frac{1}{-2\eta}(1+x^2-\eta^2),\\
    X^i&=\frac{x_i}{-\eta},\\
    X^{d+1}&=\frac{1}{-2\eta}(1-x^2+\eta^2),
\end{align}
In these coordinates, the conformal boundary $\eta\rightarrow 0^-$ of de Sitter space is $\mathbb{R}^d$.
In the $\eta\rightarrow 0^-$ limit, the degree $s-1$ pseudo-harmonic polynomials produce exactly the functions of the form $(-\eta)^{1-s}g(1,x_i,x^2)$, where $g(a_1\cdots a_{d+2})$ is degree $s-1$ polynomial. As discussed before, these functions form $E_{s,0}$.

We can also go to the ``compact'' case, where $E_{s,0}$ is viewed functions on $S^d$. It is spanned by spherical harmonics $Y_{l m_1,...}(\Omega)$ with $0 \le l \le s-1$.
The degree $s-1$ pseudo-harmonic polynomials in $\mathbb{M}^{1,d+1}$, correspond to degree $\le s-1$ harmonic polynomials on $\mathbb{R}^{d+1}$, and thus degree $\le s-1$ spherical harmonics on $S^d$. This can be seen easily by noting that if we take $X^0$ to be 1, we can go from pseudo-harmonic polynomials in $\mathbb{M}^{1,d+1}$ to those in $\mathbb{R}^{d+1}$, and adding $X^0$ we go the other way round.

On $S^d$, a function in $C_{1-s,0}$ can be expanded
\begin{equation}
\label{expansion-sphere}
    \psi(\Omega)=\sum_{l,m}\psi_{lm} Y_{l,m}(\Omega).
\end{equation}
An $SO(1,d+1)$ invariant positive definite inner product defined on the subspace $V_{s,0}=C_{1-s,0}/E_{s,0}$ is \cite{Sun:2021thf}
\begin{equation}
\label{bdy-product}
    (\psi,\phi)=\sum_{l\ge s}\sum_m \frac{\Gamma(d+s+l-1)}{\Gamma(l+1-s)}\psi_{lm}^*\phi_{lm}.
\end{equation}
We note that a non-zero function in $V_{s,0}$ is an equivalence class of functions, which have the same $\psi_{lm}$ with $l\ge s$, but can also have modes with $l \le s-1$. The generators of $SO(1,d+1)$  relate spherical harmonics of different degrees, and spherical harmonics with degree $l \ge s$ are not closed under the action of $SO(1,d+1)$. The inner product (\ref{bdy-product}) neglects the $l \le s-1$ part. If we included also the modes with $l\le s-1$ in (\ref{bdy-product}), the inner product would not be positive definite.

\subsection{Relation to $SL(2,\mathbb{C})$}

When $d=2$, for $t\ge 1$, $C_{1-t,s},\, C_{d+t-1,s}$ are irreducible. $F'_{s,t}=C_{1-t,s}$ and $D_{s,t}=0$. For $t=0$, which is the relevant case, when viewed as representations of $SL(2,\mathbb{C})$
\be
C_{1-t,s}= C_{d+t-1,s}=C_{1,s}\cong D_{(s,-s)}\oplus D_{(-s,s)},
\ee
where $D_{(s,-s)}$ is the notation used in section \ref{sl2}.
The arrow $d^s$ and $d'^s$ each split into two arrows, producing diagram (\ref{sl2diagram}). Thus $E_{s,0}=E_{(s,s)}$, and $V_{s,0}=D_{(s,s)}/E_{(s,s)}$, note that we use brackets in the subscript to denote representations discussed in section \ref{sl2}.

This can be checked directly. In two dimensions, the two complex null vectors are $z=(1,i)$ and $z=(1,-i)$,  $z\cdot \partial_x$ is proportional to 
$\partial$ or $\bar \partial$. $D_z\cdot\partial_x$ is proportional to $\bar\partial$ or $\partial$.

\section{Exceptional scalar in de Sitter space}\label{3}
In this section, we go to $d+1$ dimensional de Sitter space $dS_{d+1}$. We will see a similar story of ``modulo some polynomials'' to get a unitary representation in de Sitter space. 

$d+1$ dimensional de Sitter space arises naturally from a particular slicing of $d+2$ dimensional Minkowski space $\mathbb{M}^{1,d+1}$ \cite{deBoer:2003vf}. We write the metric of Minkowski space as
\be
ds^2=-dX_0^2+d X_1^2+\cdots+d X_{d+1}^2.
\ee 
Outside the lightcone
\be
-X_0^2+X_1^2+\cdots X_{d+1}^2=r^2>0,
\ee
we can write the metric as
\be
ds^2=dr^2+r^2 ds^2_{dS_{d+1}},
\ee
where
\be
ds^2_{dS_{d+1}}=-d\tau^2+\cosh^2\tau d\Omega_d^2
\ee
is the global coordinates on $dS_{d+1}$. $d\Omega_d^2$ is the metric on $d$ dimensional sphere $S^d$.

We intend to show that exceptional tachyon scalar
\begin{equation}
    S=\int d^{d+1} x\sqrt{-g}\left(-\frac{1}{2}(\partial\phi)^2-\frac{m^2}{2}\phi^2\right),
\end{equation}
with mass squared
\begin{equation}
    m^2=-(s-1)(s-1+d),\; s\in\mathbb{Z}_+.
\end{equation}
modulo a suitable ``shift symmetry'' \cite{Bonifacio:2018zex}, carries exactly the representation of $V_{s,0}$ mentioned in section \ref{rep}. For $d=2,\,s=2$, the representation is $D_{(2,2)}/E_{(2,2)}$, where supertranslation Goldstone lies.

Shift symmetry means
\begin{equation}
\label{shift-symmetry}
    \delta \phi = S_{a_1,... a_{s-1}} X^{a_1}... X^{a_{s-1}},
\end{equation}
where $X^i$ are coordinates of the embedding Minkowski space $\mathbb{M}^{1,d+1}$, and $S_{a_1,... a_{s-1}}$ is symmetric and traceless. In fact $S_{a_1,... a_{s-1}} X^{a_1}... X^{a_{s-1}}$ are degree $s-1$ pseudo-harmonic polynomial, and they satisfy the equation of motion. When $s=1$, the mass squared is 0, which is minimally coupled massless field. The shift symmetry corresponds to a constant shift $\delta \phi=a$.

These solutions should be modded out is similar to the case of a massless scalar field on a sphere, see, for example \cite{Folacci:1992xc}. Intuitively, when the mass squared is equal to the  eigenvalue of the Laplace operator on sphere, there is divergence in the partition function
\begin{equation}
    Z=\det(\Delta-m^2)^{-1/2}.
\end{equation}
Quotienting out these modes can remove the infinity of the partition function.

\subsection{Analysis near conformal boundary}
Exceptional de Sitter scalars are thoroughly studied in, for example \cite{Epstein:2014jaa}. Here we focus on the behavior of the theory near conformal boundary, in the spirit of dS/CFT correspondence \cite{Strominger:2001pn,Sun:2021thf}. The dS/CFT correspondence provides a more convenient way to quantize fields in de Sitter space, and also links more directly to the representation theory discussed in the previous section.

We use Poincaré coordinates of $d+1$ dimensional de Sitter space, 
\begin{equation}
    ds^2=\frac{1}{\eta^2}(-d \eta^2+\sum_i dx_i^2).
\end{equation}
The on shell solution near future infinity $\eta\rightarrow 0^-$ has asymptotics
\begin{equation}
    \phi(x,\eta)= (-\eta)^{-(s-1)} \phi_1(x)+\cdots+(-\eta)^{d+s-1}\phi_2(x)+(-\eta)^{d+s-1}\ln(-\eta)\phi'_2(x)+\cdots.
\end{equation}
Since $s$ is integral, there is $\ln (-\eta)$ term.
From the general formula
\begin{equation}
  \phi(x,\eta)= (-\eta)^{\Delta} O_1(x)+(-\eta)^{d-\Delta}O_2(x),
\end{equation}
we read the conformal dimension of the boundary operator $ O_1(x)$
\begin{equation}
\Delta=1-s.
\end{equation}

Suppose that we have a CFT on the conformal boundary $\mathbb{R}^d$, which has a ground state $\ket{0}$ and operators $O_1(x)$ acting on it generate the states
\be
O_1(x)\ket{0}=\ket{x}.
\ee
In general, it is equivalent to use either $O_1(x)$ or $O_2(x)$ to generate the states. In our case, using $O_1(x)$ is much more covenient.
The action of isometries of de Sitter space
\begin{align}
    P_i&=-\partial_i, \\
    D&=-(x\cdot\partial_x+\eta\partial_\eta),\\
    M_{ij}&=x_i\partial_j-x_j\partial_i,\\
    K_i&=(x^2-\eta^2)\partial_i-2x_i(x\cdot\partial_x+\eta\partial_\eta),
\end{align}
on $\phi(\eta,x)$ as Lie derivative
\be
[P_i,\phi(\eta,x)]\equiv-\mathcal{L}_{P_i} \phi(\eta,x),
\ee
descends to $O_1(x)$ (after taking $\eta\rightarrow0^-$ limit) as in (\ref{xp-action})-(\ref{xk-action}).

As in section \ref{rep}, we use smeared states
\be
\ket{\psi}=\int d^d x \psi(x)O_1(x)\ket{0},
\ee
Thus we see $\psi(x)$ lie in space $C_{\bar\Delta,0}=C_{d+s-1,0}$.

We have shown in section \ref{rep} that the shift symmetry corresponds to polynomials which form $E_{s,0}$. The bilinear form on $C_{1-s,0}\times C_{d+s-1,0}$ is defined by 
\be
(f(x),g(x))=\int d^d x \bar f(x) g(x).
\ee
$E_{s,0}$ and $F_{s,0}$ are orthogonal with respect to this bilinear form. Thus $\psi(x)$ should be orthogonal to $E_{s,0}$ and lies in $F_{s,0}\cong V_{s,0}$. We see that through dS/CFT correspondence, the one particle states of exceptional tachyon with shift symmetry modded out, the states $\ket{\psi}$, form $V_{s,0}$.

In \cite{Bonifacio:2018zex}, the authors noted that these scalars with shift symmetries are connected with partially massless fields with spin $s$ and depth $0$. The shift symmetry comes from the ambiguity of gauge parameters of partially massless fields. Furthermore, the partially massless fields and shift symmetric scalars can be seen as the split of massive representation as a partially massless one and a scalar
\begin{equation}
   (\Delta,s)_{\Delta\rightarrow d-1} \rightarrow (d-1,s)\oplus (s+d-1,0).
\end{equation}
Now $(d-1,s)$ is the spin $s$ and depth $0$ partially massless field, which corresponds to $D_{s,0}$ using the notation in section \ref{rep}, and the remaining $(s+d-1,0)$ is the exceptional scalar discussed in this section. In other words
\begin{equation}
\label{c/d}
    V_{s,0}\cong C_{d-1,s}/D_{s,0}.
\end{equation}

There are also spin $t$ fields with shift symmetry in de Sitter space in \cite{Bonifacio:2018zex}. With shift symmetry modded out, they are described by representation $V_{s,t}$. However, they are not unitary.

\subsection{Boundary primary operators}
Suppose operator $O(z,\bar z)$ is dual to exceptional scalar field in three dimensional de Sitter space with parameter $s$. We might try to write down action
\be
S\propto\int d^2 z O(z,\bar z)\square^s O(z,\bar z)=4^s\int d^2 z \partial^s O(z,\bar z)\bar \partial^s O(z,\bar z).
\ee
One need to quotient out the polynomials of $z,\bar z$, each with degree at most $s-1$. The conformal primary operators are $\partial^s O$ and $\bar \partial^s O$, similar to the story in section \ref{2}. 

In higher dimensions,
suppose we have a scalar theory on the boundary
\be\label{boxk}
S\propto \int d^d x O(x)\square^k O(x).
\ee
To match the conformal dimension $\Delta=1-s$ of exceptional scalar, we must have 
\be\label{k=d/2}
k=\frac{d}{2}+s-1.
\ee
The theory is local only when $d$ is an even integer.
Now the primary operators are $(z\cdot \partial)^s O(x)$, where $z$ is a complex null vector. In more familiar notation, they are traceless symmetric tensors $\partial_{\mu_1}\cdots\partial_{\mu_s} O(x)$.  They have spin $s$ and dimension $1$. 
The two point function of $O(x)$ is 
\be
\label{ooln}
\langle O(x)O(0) \rangle\propto | x |^{2(s-1)} \ln |x|^2,
\ee
from which we can calculate
\be
\label{zxs}
\langle (z\cdot \partial)^s O(x)O(0) \rangle\propto \frac{(z\cdot x)^s}{x^2},
\ee
and two point function of $ (z\cdot\partial)^sO(x)$
\be
\label{s1}
\langle (z\cdot\partial)^sO(x)(z'\cdot\partial)^s O(0) \rangle\propto \frac{\left(2 (x\cdot z)( x\cdot z')-(z\cdot z') x^2\right)^s}{(x^2)^{1+s}}.
\ee
which is the standard result of the two point function of conformal primary spin $s$ dimension $\Delta=1$ operators \cite{Costa:2011mg}.
One way to obtain (\ref{s1}) from (\ref{zxs}) is to divide $(z'\cdot\partial)^s$ into those acting on $(z\cdot x)^s$ and those acting on $\frac{1}{x^2}$. By summing these up, we can get an expanded version of the right hand side of (\ref{s1}) with an extra coefficient $s!$.


According to \cite{Bekaert:2013zya,Brust:2016gjy,Brust:2016zns}, the boundary dual of higher spin partially massless fields in (A)dS is a $\square^k$ scalar field theory, but the two point function is analytic, i.e., discard the log in (\ref{ooln}), and leave only the polynomial part. It would be interesting to further investigate what bulk high spin theory does the full boundary theory (\ref{boxk}) with log in its two point function corresponds to.

\subsection{Going to global coordinates}
In de Sitter space, the Klein-Gordon product 
\begin{equation}
\label{klein-gordon}
    (\psi,\phi)=i \int_\Sigma d^d x(\psi^*\partial_\mu\phi-\phi\partial_\mu \psi^*)
\end{equation}
defines a positive definite inner product between positive frequency modes.
We now check that on one-particles states, this coincides with the inner product  of representation $V_{s,0}$, which is given by (\ref{bdy-product}).

Using global coordinates, we expand $\phi$ as
\begin{equation}
    \phi=\sum_{l,m}a_{lm}\phi_{l}(t) Y_{lm}(\Omega)+a^\dagger_{lm}\phi_{l}^*(t)Y_{lm}^*(\Omega),
\end{equation}
where $\phi_{l}$ are normalized according to (\ref{klein-gordon})
\begin{equation}
\label{normalization}
    (\phi_{l},\phi_{l'})=\delta_{ll'}.
\end{equation}
Upon quantization, we have
\begin{equation}
    [a_{lm},a^\dagger_{l'm'}]=\delta_{ll'}\delta_{mm'},
\end{equation}
and a vacuum $\ket{0}$ which is annihilated by $a_{lm}$. This means $\phi_{l}(t)$ are positive frequency modes. The physical Hilbert space is built by acting 
$a_{lm}^\dagger$ on $\ket{0}$.

In de Sitter space, there is a conformal class of such choice of vacua \cite{Bousso:2001mw}. We choose the Bunch-Davis vacuum, which is the analytic continuation of Euclidean vacuum. This means \cite{Epstein:2014jaa}
\begin{equation}
        \phi_{l}(t)=\left(\frac{1}{2}\Gamma(l-(s-1))\Gamma(d+s-1+l)\right)^{1/2}(\cosh t)^{-(d-1)/2}P^{-l-(d-1)/2}_{s-1+(d-1)/2}(i\sinh t).
\end{equation}
$P^a_b(t)$ is Legendre function. The coefficient is to make sure $\phi_{l}$ is normalized according to (\ref{normalization}).
One can see that if $s\in\mathbb{Z}_+$ and $l\le s-1$ there is pole in the coefficient. On the conformal boundary $S^d$, these are spherical harmonics with $l\le s-1$, and form representation $E_{s,0}$ of $SO(1,d+1)$ as discussed in section \ref{rep}. We attempt to mod out this modes, and that the physical modes only contain $l\ge s$
\begin{equation}
    \phi_{\mathrm{phy}}=\sum_{l\ge s,m}a_{lm}\phi_{l}(t) Y_{lm}(\Omega)+a^\dagger_{lm}\phi_{l}^*(t)Y_{lm}^*(\Omega).
\end{equation}
The asymptotic behavior of $\phi_{l}(t)$ near $t\rightarrow\infty$ is 
\begin{equation}
    \phi_{l}(t)\sim (\cosh t) ^{-s} f(d,s)\sqrt{\frac{\Gamma(l-(s-1))}{\Gamma(d+(s-1)+l)}}\; ,
\end{equation}
where $f(d,s)$ is independent of $l$. Now the Klein-Gordon product (\ref{normalization}) evaluated on the timelike boundary match the inner product of unitary representation $V_{s,0}$ in (\ref{bdy-product}), up to a factor that is independent of $l$ and can be chosen arbitrarily. Similar to the discussion below (\ref{bdy-product}), these physical modes are not invariant under $SO(1,d+1)$. We need to discard the modes with $l\le s-1$ generated through the $SO(1,d+1)$ action on physical modes.

\section{Discussions}
In this note, we analyzed certain integer weight representations of $SO(1,d+1)$, and point out their physical realizations as supertranslation Goldstones, exceptional scalar tachyons in de Sitter space and certain $\square^k$ theories in Euclidean space. To get a physical theory, i.e., having the symmetry group unitarily represented, we must quotient out certain polynomial ``gauge'' symmetries.

The fact that the free theory of supertranslation Goldstone does not respect full Virasoro symmetry may seem disturbing.  Superrotation, celestial Virasoro symmetry and subleading gravitational soft theorem are related  \cite{Cachazo:2014fwa,Kapec:2014opa}, or even $w_{1+\infty}$ symmetry \cite{Guevara:2021abz,Strominger:2021mtt} when considering more subleading contributions, but how much does this Virasoro symmetry can tell about a realistic celestial model remains unknown.

It would be interesting to consider interacting models of supertranslation Goldstone. It would also be interesting to study other currents and Goldstones corresponding to subleading soft theorems of gravitons and photons using representation theory. However, for subleading soft theorems, the Goldstone modes have spin, and cannot form a unitary representation of $SL(2,\mathbb{C})$. The fact that subleading current or Goldstone alone do not form a unitary representation of $SL(2,\mathbb{C})$ may hint that inevitably we must take more subleading terms into consideration, since the bulk theory is unitary. This is plausible, because the action of bulk translation operator on celestial conformal primary operators changes the conformal dimension \cite{Stieberger:2018onx}. 

As mentioned in section \ref{3}, the boundary operators corresponding to exceptional scalars in de Sitter space are scalar operators which have logarithms in their two point functions. The effective action is given by a $\square^k$ theory. But the $\square^k$ theories has many more primaries. It would be interesting to further investigate whether there exist suitable higher spin theories in de Sitter space (which contain the exceptional scalars) that are holographic dual to them.

\section*{Acknowledgments}
We thank Bin Chen, Leiko Liu, Yufan Zheng for useful discussions. The work is in part supported by NSFC Grant  No. 11735001.

\bibliographystyle{utphys}
\bibliography{ref}

\providecommand{\href}[2]{#2}\begingroup\raggedright\begin{thebibliography}{10}

\bibitem{Pasterski:2016qvg}
S.~Pasterski, S.-H. Shao, and A.~Strominger, ``{Flat Space Amplitudes and
  Conformal Symmetry of the Celestial Sphere},''
  \href{http://dx.doi.org/10.1103/PhysRevD.96.065026}{{\em Phys. Rev. D}
  {\bfseries 96} no.~6, (2017) 065026},
  \href{http://arxiv.org/abs/1701.00049}{{\ttfamily arXiv:1701.00049
  [hep-th]}}.

\bibitem{Pasterski:2017kqt}
S.~Pasterski and S.-H. Shao, ``{Conformal basis for flat space amplitudes},''
  \href{http://dx.doi.org/10.1103/PhysRevD.96.065022}{{\em Phys. Rev. D}
  {\bfseries 96} no.~6, (2017) 065022},
  \href{http://arxiv.org/abs/1705.01027}{{\ttfamily arXiv:1705.01027
  [hep-th]}}.

\bibitem{Pasterski:2017ylz}
S.~Pasterski, S.-H. Shao, and A.~Strominger, ``{Gluon Amplitudes as 2d
  Conformal Correlators},''
  \href{http://dx.doi.org/10.1103/PhysRevD.96.085006}{{\em Phys. Rev. D}
  {\bfseries 96} no.~8, (2017) 085006},
  \href{http://arxiv.org/abs/1706.03917}{{\ttfamily arXiv:1706.03917
  [hep-th]}}.

\bibitem{Strominger:2017zoo}
A.~Strominger, ``{Lectures on the Infrared Structure of Gravity and Gauge
  Theory},'' \href{http://arxiv.org/abs/1703.05448}{{\ttfamily arXiv:1703.05448
  [hep-th]}}.

\bibitem{Raclariu:2021zjz}
A.-M. Raclariu, ``{Lectures on Celestial Holography},''
  \href{http://arxiv.org/abs/2107.02075}{{\ttfamily arXiv:2107.02075
  [hep-th]}}.

\bibitem{Pasterski:2021rjz}
S.~Pasterski, ``{Lectures on celestial amplitudes},''
  \href{http://dx.doi.org/10.1140/epjc/s10052-021-09846-7}{{\em Eur. Phys. J.
  C} {\bfseries 81} no.~12, (2021) 1062},
  \href{http://arxiv.org/abs/2108.04801}{{\ttfamily arXiv:2108.04801
  [hep-th]}}.

\bibitem{Strominger:2001pn}
A.~Strominger, ``{The dS / CFT correspondence},''
  \href{http://dx.doi.org/10.1088/1126-6708/2001/10/034}{{\em JHEP} {\bfseries
  10} (2001) 034}, \href{http://arxiv.org/abs/hep-th/0106113}{{\ttfamily
  arXiv:hep-th/0106113}}.

\bibitem{Weinberg:1965nx}
S.~Weinberg, ``{Infrared photons and gravitons},''
  \href{http://dx.doi.org/10.1103/PhysRev.140.B516}{{\em Phys. Rev.} {\bfseries
  140} (1965) B516--B524}.

\bibitem{Strominger:2013lka}
A.~Strominger, ``{Asymptotic Symmetries of Yang-Mills Theory},''
  \href{http://dx.doi.org/10.1007/JHEP07(2014)151}{{\em JHEP} {\bfseries 07}
  (2014) 151}, \href{http://arxiv.org/abs/1308.0589}{{\ttfamily arXiv:1308.0589
  [hep-th]}}.

\bibitem{Strominger:2013jfa}
A.~Strominger, ``{On BMS Invariance of Gravitational Scattering},''
  \href{http://dx.doi.org/10.1007/JHEP07(2014)152}{{\em JHEP} {\bfseries 07}
  (2014) 152}, \href{http://arxiv.org/abs/1312.2229}{{\ttfamily arXiv:1312.2229
  [hep-th]}}.

\bibitem{He:2014laa}
T.~He, V.~Lysov, P.~Mitra, and A.~Strominger, ``{BMS supertranslations and
  Weinberg\textquoteright{}s soft graviton theorem},''
  \href{http://dx.doi.org/10.1007/JHEP05(2015)151}{{\em JHEP} {\bfseries 05}
  (2015) 151}, \href{http://arxiv.org/abs/1401.7026}{{\ttfamily arXiv:1401.7026
  [hep-th]}}.

\bibitem{He:2014cra}
T.~He, P.~Mitra, A.~P. Porfyriadis, and A.~Strominger, ``{New Symmetries of
  Massless QED},'' \href{http://dx.doi.org/10.1007/JHEP10(2014)112}{{\em JHEP}
  {\bfseries 10} (2014) 112}, \href{http://arxiv.org/abs/1407.3789}{{\ttfamily
  arXiv:1407.3789 [hep-th]}}.

\bibitem{Donnay:2018neh}
L.~Donnay, A.~Puhm, and A.~Strominger, ``{Conformally Soft Photons and
  Gravitons},'' \href{http://dx.doi.org/10.1007/JHEP01(2019)184}{{\em JHEP}
  {\bfseries 01} (2019) 184}, \href{http://arxiv.org/abs/1810.05219}{{\ttfamily
  arXiv:1810.05219 [hep-th]}}.

\bibitem{Donnay:2020guq}
L.~Donnay, S.~Pasterski, and A.~Puhm, ``{Asymptotic Symmetries and Celestial
  CFT},'' \href{http://dx.doi.org/10.1007/JHEP09(2020)176}{{\em JHEP}
  {\bfseries 09} (2020) 176}, \href{http://arxiv.org/abs/2005.08990}{{\ttfamily
  arXiv:2005.08990 [hep-th]}}.

\bibitem{Pasterski:2021fjn}
S.~Pasterski, A.~Puhm, and E.~Trevisani, ``{Celestial diamonds: conformal
  multiplets in celestial CFT},''
  \href{http://dx.doi.org/10.1007/JHEP11(2021)072}{{\em JHEP} {\bfseries 11}
  (2021) 072}, \href{http://arxiv.org/abs/2105.03516}{{\ttfamily
  arXiv:2105.03516 [hep-th]}}.

\bibitem{Pasterski:2021dqe}
S.~Pasterski, A.~Puhm, and E.~Trevisani, ``{Revisiting the conformally soft
  sector with celestial diamonds},''
  \href{http://dx.doi.org/10.1007/JHEP11(2021)143}{{\em JHEP} {\bfseries 11}
  (2021) 143}, \href{http://arxiv.org/abs/2105.09792}{{\ttfamily
  arXiv:2105.09792 [hep-th]}}.

\bibitem{Penedones:2015aga}
J.~a. Penedones, E.~Trevisani, and M.~Yamazaki, ``{Recursion Relations for
  Conformal Blocks},'' \href{http://dx.doi.org/10.1007/JHEP09(2016)070}{{\em
  JHEP} {\bfseries 09} (2016) 070},
  \href{http://arxiv.org/abs/1509.00428}{{\ttfamily arXiv:1509.00428
  [hep-th]}}.

\bibitem{Bros:2010wa}
J.~Bros, H.~Epstein, and U.~Moschella, ``{Scalar tachyons in the de Sitter
  universe},'' \href{http://dx.doi.org/10.1007/s11005-010-0406-4}{{\em Lett.
  Math. Phys.} {\bfseries 93} (2010) 203--211},
  \href{http://arxiv.org/abs/1003.1396}{{\ttfamily arXiv:1003.1396 [hep-th]}}.

\bibitem{Epstein:2014jaa}
H.~Epstein and U.~Moschella, ``{de Sitter tachyons and related topics},''
  \href{http://dx.doi.org/10.1007/s00220-015-2308-x}{{\em Commun. Math. Phys.}
  {\bfseries 336} no.~1, (2015) 381--430},
  \href{http://arxiv.org/abs/1403.3319}{{\ttfamily arXiv:1403.3319 [hep-th]}}.

\bibitem{Bonifacio:2018zex}
J.~Bonifacio, K.~Hinterbichler, A.~Joyce, and R.~A. Rosen, ``{Shift Symmetries
  in (Anti) de Sitter Space},''
  \href{http://dx.doi.org/10.1007/JHEP02(2019)178}{{\em JHEP} {\bfseries 02}
  (2019) 178}, \href{http://arxiv.org/abs/1812.08167}{{\ttfamily
  arXiv:1812.08167 [hep-th]}}.

\bibitem{Bonifacio:2021mrf}
J.~Bonifacio, K.~Hinterbichler, A.~Joyce, and D.~Roest, ``{Exceptional scalar
  theories in de Sitter space},''
  \href{http://dx.doi.org/10.1007/JHEP04(2022)128}{{\em JHEP} {\bfseries 04}
  (2022) 128}, \href{http://arxiv.org/abs/2112.12151}{{\ttfamily
  arXiv:2112.12151 [hep-th]}}.

\bibitem{Bondi:1962px}
H.~Bondi, M.~G.~J. van~der Burg, and A.~W.~K. Metzner, ``{Gravitational waves
  in general relativity. 7. Waves from axisymmetric isolated systems},''
  \href{http://dx.doi.org/10.1098/rspa.1962.0161}{{\em Proc. Roy. Soc. Lond. A}
  {\bfseries 269} (1962) 21--52}.

\bibitem{Sachs:1962wk}
R.~K. Sachs, ``{Gravitational waves in general relativity. 8. Waves in
  asymptotically flat space-times},''
  \href{http://dx.doi.org/10.1098/rspa.1962.0206}{{\em Proc. Roy. Soc. Lond. A}
  {\bfseries 270} (1962) 103--126}.

\bibitem{Himwich:2020rro}
E.~Himwich, S.~A. Narayanan, M.~Pate, N.~Paul, and A.~Strominger, ``{The Soft
  $\mathcal{S}$-Matrix in Gravity},''
  \href{http://dx.doi.org/10.1007/JHEP09(2020)129}{{\em JHEP} {\bfseries 09}
  (2020) 129}, \href{http://arxiv.org/abs/2005.13433}{{\ttfamily
  arXiv:2005.13433 [hep-th]}}.

\bibitem{Arkani-Hamed:2020gyp}
N.~Arkani-Hamed, M.~Pate, A.-M. Raclariu, and A.~Strominger, ``{Celestial
  amplitudes from UV to IR},''
  \href{http://dx.doi.org/10.1007/JHEP08(2021)062}{{\em JHEP} {\bfseries 08}
  (2021) 062}, \href{http://arxiv.org/abs/2012.04208}{{\ttfamily
  arXiv:2012.04208 [hep-th]}}.

\bibitem{Naculich:2011ry}
S.~G. Naculich and H.~J. Schnitzer, ``{Eikonal methods applied to gravitational
  scattering amplitudes},''
  \href{http://dx.doi.org/10.1007/JHEP05(2011)087}{{\em JHEP} {\bfseries 05}
  (2011) 087}, \href{http://arxiv.org/abs/1101.1524}{{\ttfamily arXiv:1101.1524
  [hep-th]}}.

\bibitem{Nguyen:2021ydb}
K.~Nguyen and J.~Salzer, ``{Celestial IR divergences and the effective action
  of supertranslation modes},''
  \href{http://dx.doi.org/10.1007/JHEP09(2021)144}{{\em JHEP} {\bfseries 09}
  (2021) 144}, \href{http://arxiv.org/abs/2105.10526}{{\ttfamily
  arXiv:2105.10526 [hep-th]}}.

\bibitem{Compere:2011ve}
G.~Compere and F.~Dehouck, ``{Relaxing the Parity Conditions of Asymptotically
  Flat Gravity},'' \href{http://dx.doi.org/10.1088/0264-9381/28/24/245016}{{\em
  Class. Quant. Grav.} {\bfseries 28} (2011) 245016},
  \href{http://arxiv.org/abs/1106.4045}{{\ttfamily arXiv:1106.4045 [hep-th]}}.
  [Erratum: Class.Quant.Grav. 30, 039501 (2013)].

\bibitem{Pate:2019mfs}
M.~Pate, A.-M. Raclariu, and A.~Strominger, ``{Conformally Soft Theorem in
  Gauge Theory},'' \href{http://dx.doi.org/10.1103/PhysRevD.100.085017}{{\em
  Phys. Rev. D} {\bfseries 100} no.~8, (2019) 085017},
  \href{http://arxiv.org/abs/1904.10831}{{\ttfamily arXiv:1904.10831
  [hep-th]}}.

\bibitem{harish1947infinite}
Harish-Chandra, ``Infinite irreducible representations of the lorentz group,''
  {\em Proceedings of the Royal Society of London. Series A. Mathematical and
  Physical Sciences} {\bfseries 189} no.~1018, (1947) 372--401.

\bibitem{bargmann1947irreducible}
V.~Bargmann, ``Irreducible unitary representations of the lorentz group,'' {\em
  Annals of Mathematics} (1947) 568--640.

\bibitem{gel1947unitary}
I.~M. Gel'fand and M.~A. Naimark, ``Unitary representations of the lorentz
  group,'' {\em Izvestiya Rossiiskoi Akademii Nauk. Seriya Matematicheskaya}
  {\bfseries 11} no.~5, (1947) 411--504.

\bibitem{gelfand1966generalized}
I.~M. Gelfand, M.~I. Graev, and N.~I. Vilenkin, {\em Generalized
  Functions-Volume 5. Integral Geometry and Representation Theory}.
\newblock Academic Press, 1966.

\bibitem{wiese1996classification}
K.~J. Wiese, ``Classification of perturbations for membranes with bending
  rigidity,'' {\em Physics Letters B} {\bfseries 387} no.~1, (1996) 57--63.

\bibitem{Karananas:2015ioa}
G.~K. Karananas and A.~Monin, ``{Weyl vs. Conformal},''
  \href{http://dx.doi.org/10.1016/j.physletb.2016.04.001}{{\em Phys. Lett. B}
  {\bfseries 757} (2016) 257--260},
  \href{http://arxiv.org/abs/1510.08042}{{\ttfamily arXiv:1510.08042
  [hep-th]}}.

\bibitem{Brust:2016gjy}
C.~Brust and K.~Hinterbichler, ``{Free \ensuremath{\square}$^{k}$ scalar
  conformal field theory},''
  \href{http://dx.doi.org/10.1007/JHEP02(2017)066}{{\em JHEP} {\bfseries 02}
  (2017) 066}, \href{http://arxiv.org/abs/1607.07439}{{\ttfamily
  arXiv:1607.07439 [hep-th]}}.

\bibitem{paneitz2008quartic}
S.~M. Paneitz {\em et~al.}, ``A quartic conformally covariant differential
  operator for arbitrary pseudo-riemannian manifolds (summary),'' {\em SIGMA.
  Symmetry, Integrability and Geometry: Methods and Applications} {\bfseries 4}
  (2008) 036.

\bibitem{Dobrev:1977qv}
V.~K. Dobrev, G.~Mack, V.~B. Petkova, S.~G. Petrova, and I.~T. Todorov,
  \href{http://dx.doi.org/10.1007/BFb0009678}{{\em {Harmonic Analysis on the
  n-Dimensional Lorentz Group and Its Application to Conformal Quantum Field
  Theory}}}, vol.~63.
\newblock 1977.

\bibitem{Sun:2021thf}
Z.~Sun, ``{A note on the representations of $\text{SO}(1,d+1)$},''
  \href{http://arxiv.org/abs/2111.04591}{{\ttfamily arXiv:2111.04591
  [hep-th]}}.

\bibitem{Deser:2001pe}
S.~Deser and A.~Waldron, ``{Gauge invariances and phases of massive higher
  spins in (A)dS},''
  \href{http://dx.doi.org/10.1103/PhysRevLett.87.031601}{{\em Phys. Rev. Lett.}
  {\bfseries 87} (2001) 031601},
  \href{http://arxiv.org/abs/hep-th/0102166}{{\ttfamily arXiv:hep-th/0102166}}.

\bibitem{Deser:2001us}
S.~Deser and A.~Waldron, ``{Partial masslessness of higher spins in (A)dS},''
  \href{http://dx.doi.org/10.1016/S0550-3213(01)00212-7}{{\em Nucl. Phys. B}
  {\bfseries 607} (2001) 577--604},
  \href{http://arxiv.org/abs/hep-th/0103198}{{\ttfamily arXiv:hep-th/0103198}}.

\bibitem{deBoer:2003vf}
J.~de~Boer and S.~N. Solodukhin, ``{A Holographic reduction of Minkowski
  space-time},'' \href{http://dx.doi.org/10.1016/S0550-3213(03)00494-2}{{\em
  Nucl. Phys. B} {\bfseries 665} (2003) 545--593},
  \href{http://arxiv.org/abs/hep-th/0303006}{{\ttfamily arXiv:hep-th/0303006}}.

\bibitem{Folacci:1992xc}
A.~Folacci, ``{BRST quantization of the massless minimally coupled scalar field
  in de Sitter space: Zero modes, euclideanization and quantization},''
  \href{http://dx.doi.org/10.1103/PhysRevD.46.2553}{{\em Phys. Rev. D}
  {\bfseries 46} (1992) 2553--2559},
  \href{http://arxiv.org/abs/0911.2064}{{\ttfamily arXiv:0911.2064 [gr-qc]}}.

\bibitem{Costa:2011mg}
M.~S. Costa, J.~Penedones, D.~Poland, and S.~Rychkov, ``{Spinning Conformal
  Correlators},'' \href{http://dx.doi.org/10.1007/JHEP11(2011)071}{{\em JHEP}
  {\bfseries 11} (2011) 071}, \href{http://arxiv.org/abs/1107.3554}{{\ttfamily
  arXiv:1107.3554 [hep-th]}}.

\bibitem{Bekaert:2013zya}
X.~Bekaert and M.~Grigoriev, ``{Higher order singletons, partially massless
  fields and their boundary values in the ambient approach},''
  \href{http://dx.doi.org/10.1016/j.nuclphysb.2013.08.015}{{\em Nucl. Phys. B}
  {\bfseries 876} (2013) 667--714},
  \href{http://arxiv.org/abs/1305.0162}{{\ttfamily arXiv:1305.0162 [hep-th]}}.

\bibitem{Brust:2016zns}
C.~Brust and K.~Hinterbichler, ``{Partially Massless Higher-Spin Theory},''
  \href{http://dx.doi.org/10.1007/JHEP02(2017)086}{{\em JHEP} {\bfseries 02}
  (2017) 086}, \href{http://arxiv.org/abs/1610.08510}{{\ttfamily
  arXiv:1610.08510 [hep-th]}}.

\bibitem{Bousso:2001mw}
R.~Bousso, A.~Maloney, and A.~Strominger, ``{Conformal vacua and entropy in de
  Sitter space},'' \href{http://dx.doi.org/10.1103/PhysRevD.65.104039}{{\em
  Phys. Rev. D} {\bfseries 65} (2002) 104039},
  \href{http://arxiv.org/abs/hep-th/0112218}{{\ttfamily arXiv:hep-th/0112218}}.

\bibitem{Cachazo:2014fwa}
F.~Cachazo and A.~Strominger, ``{Evidence for a New Soft Graviton Theorem},''
  \href{http://arxiv.org/abs/1404.4091}{{\ttfamily arXiv:1404.4091 [hep-th]}}.

\bibitem{Kapec:2014opa}
D.~Kapec, V.~Lysov, S.~Pasterski, and A.~Strominger, ``{Semiclassical Virasoro
  symmetry of the quantum gravity $ \mathcal{S}$-matrix},''
  \href{http://dx.doi.org/10.1007/JHEP08(2014)058}{{\em JHEP} {\bfseries 08}
  (2014) 058}, \href{http://arxiv.org/abs/1406.3312}{{\ttfamily arXiv:1406.3312
  [hep-th]}}.

\bibitem{Guevara:2021abz}
A.~Guevara, E.~Himwich, M.~Pate, and A.~Strominger, ``{Holographic symmetry
  algebras for gauge theory and gravity},''
  \href{http://dx.doi.org/10.1007/JHEP11(2021)152}{{\em JHEP} {\bfseries 11}
  (2021) 152}, \href{http://arxiv.org/abs/2103.03961}{{\ttfamily
  arXiv:2103.03961 [hep-th]}}.

\bibitem{Strominger:2021mtt}
A.~Strominger, ``{$w_{1+\infty}$ Algebra and the Celestial Sphere: Infinite
  Towers of Soft Graviton, Photon, and Gluon Symmetries},''
  \href{http://dx.doi.org/10.1103/PhysRevLett.127.221601}{{\em Phys. Rev.
  Lett.} {\bfseries 127} no.~22, (2021) 221601}.

\bibitem{Stieberger:2018onx}
S.~Stieberger and T.~R. Taylor, ``{Symmetries of Celestial Amplitudes},''
  \href{http://dx.doi.org/10.1016/j.physletb.2019.03.063}{{\em Phys. Lett. B}
  {\bfseries 793} (2019) 141--143},
  \href{http://arxiv.org/abs/1812.01080}{{\ttfamily arXiv:1812.01080
  [hep-th]}}.

\end{thebibliography}\endgroup

\end{document}